\documentclass[]{article}
\usepackage{emulateapj}
\usepackage{graphicx}
\usepackage{multirow}

\advance \voffset by -0.5cm\relax

\usepackage{color}

\def\msun{{\rm ~M}_{\odot}}
\def\rsun{{\rm ~R}_{\odot}}

\def\mpy{{\rm ~M}_{\odot} {\rm ~yr}^{-1}}
\def\zsun{{\rm ~Z}_{\odot}}

\begin{document}

\title{The effect of metallicity on the detection prospects for gravitational waves}

 \author{Krzysztof Belczynski\altaffilmark{1,2},
         Michal Dominik\altaffilmark{2},
         Tomasz Bulik\altaffilmark{2},
         Richard O'Shaughnessy\altaffilmark{3},
         Chris Fryer \altaffilmark{1},
         Daniel E. Holz \altaffilmark{1}}

 \affil{
     $^{1}$ Los Alamos National Laboratory, P.O. Box 1663, Los Alamos, NM 87545, USA \\
     $^{2}$ Astronomical Observatory, University of Warsaw, Al.
            Ujazdowskie 4, 00-478 Warsaw, Poland\\
     $^{3}$ Department of Physics, Penn State University, 104 Davey Lab, University Park, PA 16802
 }
 
\begin{abstract}
Data from the SDSS ($\sim 300,000$ galaxies) indicates that recent star
formation (within the last 1 billion years) is bimodal: half the stars form
from gas with high amounts of metals (solar metallicity), and the other half
form with small contribution of elements heavier than Helium ($\sim 10-30\%$
solar). Theoretical studies of mass loss from the brightest stars derive
significantly higher stellar-origin BH masses ($\sim 30-80 \msun$)
than previously estimated for sub-solar compositions. We combine these
findings to estimate the probability of detecting gravitational waves
(GWs) arising from the inspiral of double compact objects. Our results show that a
low metallicity environment significantly boosts the formation of double
compact object binaries with at least one BH. In particular,
we find the GW detection rate is increased by a factor of
$20$ if the metallicity is decreased from solar (as in all previous
estimates) to a 50--50 mixture of solar and $10\%$ solar metallicity. The
current sensitivity of the two largest instruments to NS-NS binary inspirals 
(VIRGO: $\sim 9$ Mpc; LIGO: $\sim 18$) is not high enough to ensure a first 
detection. However, our results indicate that if a future instrument 
increased the sensitivity to $\sim 50-100\,\mbox{Mpc}$, a detection of
GWs would be expected within the first year of observation.
It was previously thought that NS-NS inspirals were the most
likely source for GW detection. Our results indicate that BH-BH binaries are
$\sim 25$-times more likely sources than NS-NS systems and that we are on the 
cusp of GW detection.
\end{abstract}


\section{Introduction}

Gravitational waves (GWs) are a consequence of Einstein's (1918) theory of general
relativity. The first indirect evidence for the existence of GWs was presented by 
Hulse and Tylor (1974) through measurement of the orbital
decay of a double binary pulsar. After four decades of effort, we have yet
to directly detect GWs. In the past
decade, two large interferometric observatories have been built to search for
gravitational waves: LIGO and VIRGO. The first observations
have been collected, and although no positive detection has been made, some useful
upper limits have been placed (Abbott et al. 2008, 2009a). Both observatories will 
soon undergo major upgrades aimed to increase their sensitivity roughly ten-fold. 
In this {\em Letter} we show that existing detectors are at the cusp of detecting
gravitational waves, and even modest improvements are likely to lead to the
first direct detection of gravitational waves.

We utilize the {\tt StarTrack} population synthesis code (Belczynski et al.
2002) to perform a suite
of Monte Carlo simulations of the stellar evolution of stars in environments
with a range of metallicity. Our calculations are guided by recent results from
the Sloan Digital Sky Survey observations (Panter et al. 2008), combined with 
recent estimates of mass loss rates (Belczynski et al. 2010a). Utilizing these 
latest results, we calculate a population of 2 million massive binary stars, 
tracking the ensuing formation of relativistic
binary compact objects: double neutron stars (NS-NS), double black hole binaries 
(BH-BH), and mixed systems (BH-NS). Our modeling utilizes updated stellar and 
binary physics, including results from supernova simulations (Fryer \& Kalogera 
2001) and compact object formation (Timmes et al. 1996), incorporating elaborate 
mechanisms for treating stellar interactions like mass transfer episodes (Belczynski 
et al. 2008) and tidal synchronization and circularization (Hut 1981). We put 
special emphasis on the common envelope evolution phase (Webbink 1984), which is
crucial for close double compact object formation as the attendant mass transfer 
allows for efficient hardening of the binary. This orbital contraction can be
sufficiently efficient to cause the individual stars in the binary to coalesce
and form a  single highly rotating object, thereby aborting further binary 
evolution and preventing the formation of a double compact object. Due to 
significant radial expansion, stars crossing the Hertzsprung gap (HG) very 
frequently initiate a common envelope phase. HG stars do not have a clear 
entropy jump at the core-envelope transition (Ivanova \& Taam 2004); if such a 
star overflows its Roche lobe and initiates a common envelope phase, the 
inspiral is expected to lead to a coalescence (Taam \& Sandquist 2000). In 
particular, it has been estimated that for a solar metallicity environment (e.g., 
our Galaxy), properly accounting for the HG gap may lead to a reduction in the 
merger rates of BH-BH binaries by $\sim 2-3$ orders of magnitude (Belczynski et 
al. 2007). The details of the common envelope phase are not yet fully understood, 
and thus in what follows we consider two models, one which does not take into 
account the suppression (optimistic model: A), and one that assumes the maximum 
suppression (pessimistic model: B).

\section{Results}

The merger and detection rates are presented in Tables~1 and~2, respectively. 
The merger rates are calculated for a Milky Way type galaxy, and therefore 
correspond to the rates within a limited volume (or a fixed star forming mass). 
The detection rates are obtained via extrapolation of the Galactic rates to an 
appropriate distance for gravitational wave observatories. The farthest 
detectable distance for a given double compact object chirp mass is given by 
$d=d_{\rm 0,nsns}(M_{\rm c,dco}/M_{\rm c,nsns})^{5/6}$, where $M_{\rm
c,dco}$ is the chirp mass of a given double compact object, $M_{\rm
c,nsns}=1.2 \msun$ is a typical chirp mass of a NS-NS binary, and $d_{\rm
0,nsns}$ is the effective distance horizon for detection of a NS-NS inspiral
(Belczynski et al. 2007). VIRGO and LIGO have currently reached 
$d_{\rm 0,nsns}=9,\ 18$ Mpc, respectively. This distance is sky and angle 
averaged\footnote{Our range is precisely the radius of a sphere with the 
same average detection volume (i.e.,$ d = ([<V T>/T]3/4pi)^{1/3}$). For  
a single detector with stationary noise, this average corresponds to a sky and 
angle averaged single-detector range $d_0$; this value is quoted in the text.  
Other authors, such as Abadie et al. (2010), describe network sensitivity
using the ``horizon distance'' $d_{\rm h}\approx 2.26 d_0$, the range to which 
that detector can see a single optimally oriented binary.}.
The distance limit is significantly 
larger for higher chirp mass double compact objects: BH-NS and BH-BH. The 
predicted chirp mass distribution (as none of the BH double compact objects are yet
observed) is presented in Figure~1.

For solar metallicity the rates show a mixture of NS-NS, BH-NS and BH-BH 
binaries in the optimistic case (model A), while BH-NS and BH-BH binaries
virtually disappear from the double compact object population in the pessimistic 
case (model B), and even rates for NS-NS are lower.  
Although NS-NS mergers dominate the rates by an order of magnitude in model 
A, the detection rates are dominated by BH-BH binaries. This is because the
heavier BH-BH binaries ($M_{\rm c,dco} \sim 5 \msun$; see Fig.~1) are ``louder''
in GWs in the appropriate band (Flanagan \& Hughes 1998), and are therefore visible 
farther than NS-NS systems ($d \sim 4 d_{\rm 0,nsns}$), resulting in $\sim 60$-times
increase in sampled volume, and a similar rate increase in detection.
In model B, there are so few BH-BH systems that NS-NS inspirals strongly 
dominate both the merger and detection rates. 

The decrease of metallicity significantly affects rates for binaries containing
BHs. We find that inspirals of double compact objects 
with BHs dominate both the merger and detection rates for stellar populations 
with $10\%$ solar metallicity, independent of our choice of model. The
rate increase for BH-BH binaries is dramatic: factors of $\sim 50/3000$-times 
for merger rates and $\sim 500/2000$-times for detection rates, for model A/B. 
These drastic enhancements, although not previously noted, are readily 
understood in terms of black hole formation physics.

In model A the rate increase is primarily due to larger BH masses at low
metallicity. For these higher BH masses we expect little to no natal kick at 
BH formation, and this leads to the survival of large fractions of massive 
binaries through supernova explosions, resulting in a significantly enhanced
formation of BH-BH systems. At low metallicity the stellar wind mass loss is
inefficient, and stars lose less mass than their higher metallicity
counterparts (Vink 2008). On the  atomic level this is understood in terms of 
interaction of radiation with matter. Winds for massive stars are 
radiation driven. The heavy elements in stellar atmosphere (large cross 
sections) intercept photons more efficiently than in stars with low metal 
content and the mass outflow from the high metallicity star is higher. For 
metal poor stars the outflows are weaker and in the end stars finish their 
evolution at higher mass. On average, in low metallicity systems the BH mass 
in close BH-BH binaries is predicted to be about $\sim 15 \msun$, roughly 
double the corresponding mass at solar  metallicity. This is shown in Figure~1 
in terms of chirp mass, and has been noted in recent observational BH mass 
estimates. While BHs in our Galaxy (approximately solar metallicity) reach 
only $\sim 15 \msun$ (Casares 2007), in galaxies of low metallicity stars form more 
massive BHs, e.g., $\sim 20 \msun$ in NGC300 ($40\% \zsun$; Crowther et al.
2010) or $\sim 30 \msun$ in IC10 ($30\% \zsun$; Prestwich et al. 2007).

These heavier masses affect binary retention. One of the most disruptive processes 
in the formation of double compact object binaries is a supernova explosion, 
and any asymmetries leading to natal kicks for the newly formed NSs, and possibly
BHs. Since NSs and BHs originate from massive stars, they form on wide (weakly
bound) orbits. Hence the first supernova tends to disrupt most 
($\gtrsim 90 \%$; Belczynski et al. 2010c) progenitor binaries. The magnitude of 
natal kicks was established for Galactic single pulsars at the level of 
$200\mbox{--}300$ km s$^{-1}$ (Hobbs et al. 2005). There is growing observational evidence that,
although low mass BHs may receive small-to-moderate kicks, massive
BHs are born in the dark (without attendant energetic supernova explosions), and
without natal kicks (Mirabel \& Rodrigues 2003; Dhawan et al. 2007; Martin
et al. 2010). These observations fit with a model
assuming that kicks are produced by asymmetric ejecta.  Even if the total
kick momentum were comparable for collapse to a black hole,
the higher the black hole mass, the lower the kick velocity.
Additionally it is likely that for very high masses ($M_{\rm zams}
\gtrsim 40 \msun$) stars promptly collapse to massive BHs, with no mass 
ejection and possibly no natal kicks (Fryer \& Kalogera 2001). We have utilized
the above information and incorporated a low mass BH/small kick and
high mass BH/no kick scheme into our population synthesis
calculations. This approach was folded in with metallicity-dependent BH
formation, leading to new estimates of the corresponding BH-BH rates.
The increase in the BH-BH merger rate is due to the lack of natal kicks in
the formation of the majority of BHs at low metallicity, and thus a high binary
survival rate through collapse. The detection rates are further boosted by the
high BH-BH chirp mass, allowing detections to much farther distance.

The results for model B show a particularly dramatic increase in the merger
rates with decreasing metallicity. BH binaries, almost non-existent at solar
metallicity, become the dominant population at low metallicity.
This striking increase in the rate is the direct result of small
stellar radii at low metallicity. This trend, noted in stellar models
(Hurley et al. 2000)  and included in our calculations, is driven by the 
decreased opacity in matter 
with fewer heavy elements; the stars are less puffed up by photons emerging 
from the core. The smaller radial expansion of stars may also be caused by the 
high rotation that is usually induced by binarity or low stellar metallicity 
(Vazquez et al. 2007; de Mink et al. 2009), but the effects of rotation on 
stellar evolution are not included in our analysis.  
These smaller stellar radii at low metallicity cause some fraction of massive 
stars to bypass mass transfer early in their evolution, and thus avoid 
coalescing in the common envelope phase. These stars survive to form a 
population of close BH-BH binaries. Figure~2 demonstrates how the 
evolution of a typical BH-BH progenitor binary alters with the change of 
metallicity.

\section{Discussion}

Recent observational data indicates that about half of recent 
star formation arises in low metallicity galaxies (Panter et al. 2008). We have created a
synthetic model of the local Universe that 
assumes half of the stars are forming at solar metallicity, and the other half
are forming at $10\%$ solar. This range of metallicity has profound effects on
the detectable rate of gravitational waves from binary systems. The
predicted detection rates are presented in Table~2, as a function of
gravitational detector sensitivity. LIGO (with its sensitivity of about 18 Mpc) has
yet to detect a binary inspiral (Abbott et al. 2009b). Our optimistic
model (A), with a predicted detection rate of $5$ yr$^{-1}$ for $d_{\rm 0,nsns}=18$ 
Mpc, is thus already excluded by LIGO. Our pessimistic model (B), with a much lower 
detection rate of $0.05$ yr$^{-1}$ for $d_{\rm 0,nsns}=18$ Mpc, is consistent with 
observations. This hints that some fraction of massive binaries cannot be surviving 
the common envelope phase.

We employ the results of our pessimistic model (B) to estimate the sensitivity
at which instruments like LIGO or VIRGO would detect 1 and 10 inspirals per
year. Assuming a local density of Milky Way-size galaxies of $\rho_{\rm
gal} = 0.01$ Mpc$^{-3}$ (O'Shaughnessy et al. 2008), we extrapolate to a distance 
sufficient to contain enough star forming mass to generate the desired
merger rate (our calculated rate of mergers per Milky Way-size galaxy are 
listed in Table~1). It is found that instruments need to reach a sensitivity of 
$d_{\rm 0,nsns}=45\,\mbox{Mpc}$ and $97\,\mbox{Mpc}$ to provide $1$ and $10$ detections
per year, respectively. We also note that the detections will be most likely
dominated by BH-BH binaries ($\sim 80\%$), with a smaller
contribution from BH-NS systems ($\sim 15\%$), and a negligible contribution from
NS-NS inspirals ($\sim 5\%$).  

Our results are subject to a number of uncertainties. First, the star 
formation rates may be different from those presented in Panter et al. (2008). Stars
which formed in the remote past but are merging at present day may manifest themselves 
as an additional and significant component to the detection rates. In-depth study
of the overall cosmic star formation rate, and the contribution of mergers from
stars born at high redshift, using existing tools (Belczynski et al. 2010b) is 
underway. Second, the process of BH formation is still somewhat
uncertain. Evolutionary effects like wind mass loss rates
may alter the final BH mass, while natal kicks (or lack thereof) or the
disputed mass delineation between NS and BH formation may 
substantially affect the rates. Although we have chosen the best
available models to describe these processes, future theoretical and observational
constraints may significantly affect our estimates.
Finally, we can exclude 100\% of stars following model A
(ignoring suppression through the common envelope phase), since this produces
rates that should have already been detected by LIGO. However, mixtures of model
A and model B (e.g., 50--50 of each model) are still consistent with
observational limits. The physical interpretation of this sort of mixture model
is that massive stars begin developing core-envelope 
structure halfway through the Hertzsprung gap, and the survival through common
envelope phase depends not only on the donor type, but also on its
evolutionary state. Therefore a fraction of binaries survive the common envelope
phase.

Our results suggest a very high potential for the detection of
gravitational radiation from a binary inspiral in the near future.
Our merger rates for the most likely source to be detected,
a BH-BH binary, are significantly higher than previous estimates in the 
literature. For example, Abadie et al. (2010) estimated the realistic level
of BH-BH detection at the level of $20$ yr$^{-1}$ for the advanced LIGO,
while our realistic prediction is of the order of $\gtrsim 200$ yr$^{-1}$. 
Note that we have not included the potentially significant contribution of
dynamically formed BH-BH binaries in globular clusters (e.g., Sadowski et
al. 2008; Downing et al. 2010). Our prediction is based 
on the most recent observational and theoretical results, and employs 
a standard stellar evolutionary model. Further support for our enhanced
theoretical predictions is the existence of high mass X-ray binaries, such as 
IC10 X-1 and NGC300 X-1, that are expected to form close BH-BH binaries
on a very short timescale (set by the evolution of the massive Wolf-Rayet 
companion). These systems have formed massive BHs in low metallicity
environment (consistent with our results), but also using their nearby locations
leads to a semi-empirical estimate of merger rates that are comparable to the 
rates presented here (Bulik et al. 2008). Extending the sensitivity of existing 
instruments by factors of only a few, to about $50\mbox{--}100$ Mpc, should 
result in the first detection of gravitational waves. The lack of detections at 
increased sensitivity will put stringent constraints on the BH-BH formation (BH 
natal kicks) and/or on the star formation (metallicity) in the local Universe.

\acknowledgements
We thank Jarrod Hurley, Peter Eggleton and Ilya Mandel for useful 
discussions. KB, TB and MD acknowledge partial support from Polish MSHE 
grants N N203 302835 and N N203 511238. KB, CF, and DH acknowledge support
from LANL under contract DE-AC52-06NA25396.

\clearpage

\begin{deluxetable}{lccc}
\tablewidth{250pt}
\tablecaption{Galactic Merger Rates [Myr$^{-1}$]\tablenotemark{a}}
\tablehead{& $\zsun$ & $0.1\ \zsun$ & $\zsun$ + $0.1\ \zsun$  \\                
 Type & ($100\%$)    & ($100\%$) & ($50\%$ + $50\%$) }

\startdata
NS-NS               &  40.8 (14.4)  &  41.3 (3.3)  &  41.1 (8.9) \\
BH-NS               &  3.2  (0.01)  &  12.1 (7.0)  &   7.7 (3.5) \\
BH-BH               &  1.5 (0.002)  &  84.2 (6.1)  &  42.9 (3.1) \\
TOTAL               &  45.5 (14.4)  &  138 (16.4)  &  91.7 (15.4) \\  
\enddata
\label{numall}
\tablenotetext{a}{
Rates are calculated for a Milky Way type galaxy (10 Gyr of continuous star
formation at a rate of $3.5 \mpy$), with the assumption that all
stars have either solar metallicity or $10\%$ solar, or a 50-50 mixture of both
types of stars. The rates are presented for the optimistic model (A) where
progenitor binaries survive through the common envelope phase, while the results
in parentheses represent the pessimistic model (B), where the binaries do not
survive if the phase is initiated by a Hertzsprung gap star.  
}
\end{deluxetable}

\begin{deluxetable}{llccc}
\tablewidth{350pt}
\tablecaption{LIGO/VIRGO Detection Rates [yr$^{-1}$] \tablenotemark{a}}
\tablehead{Sensitivity & & $\zsun$ & $0.1\ \zsun$ & $\zsun$ + $0.1\ \zsun$  \\
($d_{\rm 0,nsns}$=) & Type & ($100\%$)    & ($100\%$) & ($50\%$ + $50\%$)}
\startdata
\multirow{4}{*}{18 Mpc} 
&NS-NS                                     &  0.01 (0.003)  &  0.01 (0.001)    &  0.01 (0.002)  \\
&BH-NS                                     &  0.007 (0.00002)  &  0.04 (0.02)  &  0.02 (0.01) \\
&BH-BH                                     &  0.02 (0.00005)   &  9.9 (0.1)    &   4.9 (0.05) \\
&TOTAL                                     &  0.03 (0.003)      & 10.0 (0.1)   &  5.0 (0.06) \\
&&&\\
\multirow{4}{*}{45 Mpc} 
&NS-NS                                       &  0.2 (0.05)    &  0.2 (0.01)    &  0.2 (0.03)  \\
&BH-NS                                       &  0.1 (0.0003)  &  0.5 (0.3)     &  0.3 (0.15) \\
&BH-BH                                       &  0.3 (0.0007)  &  145.4 (1.6)   &   72.8 (0.82) \\
&TOTAL                                       &  0.6 (0.05)    & 146.1 (1.9)    &   73.3 (1.0) \\
&&&\\
\multirow{4}{*}{97 Mpc}
&NS-NS                                       &  1.5 (0.5)    &  1.6 (0.1)       &  1.5 (0.3)  \\
&BH-NS                                       &  1.0 (0.003)  &  4.8 (2.9)       &  2.9 (1.5) \\
&BH-BH                                       &  2.8 (0.007)  &  1454.6 (16.4)   &  728.7 (8.2) \\
&TOTAL                                       &  5.3 (0.5)    &  1461.0 (19.5)   &  733.2 (10.0) \\
&&&\\
\multirow{4}{*}{300 Mpc} 
&NS-NS                                       &  44.3 (15.1)  &  45.9 (4.0)      &  45.1 (9.5)  \\
&BH-NS                                       &   29.7 (0.1)  &  141.9 (85.4)    &  85.8 (42.8) \\
&BH-BH                                       &  82.4 (0.21)  &  42768.0 (483.3) &  21425.2 (241.7) \\
&TOTAL                                       &  156.4 (15.2) & 42955.8 (572.7)  & 21556.0 (294.0) \\
\enddata
\label{Lmerger}
\tablenotetext{a}{Detection rates for model A (B) as a function of
sensitivity of a given instrument. 
Sensitivity is defined as the sky and angle averaged distance horizon for
detection of a double neutron star inspiral. 
The rates are given for a local Universe consisting of only solar composition
stars (unrealistically high), $0.1 \zsun$ stars (unrealistically low) and for
a 50-50 mixture of the above (realistic local Universe;~Panter et al. 2008). The sensitivity 
of $d_{\rm 0,nsns}=18\,\mbox{Mpc}$ and $300\,\mbox{Mpc}$ correspond to the current and 
advanced (2015) LIGO detector, respectively. $d_{\rm 0,nsns}=45\,\mbox{Mpc}$ and
$97\,\mbox{Mpc}$ were chosen such that our pessimistic model (B) results in 1
and 10 detections, respectively.
}
\end{deluxetable}

\begin{figure}
\includegraphics[width=0.7\columnwidth]{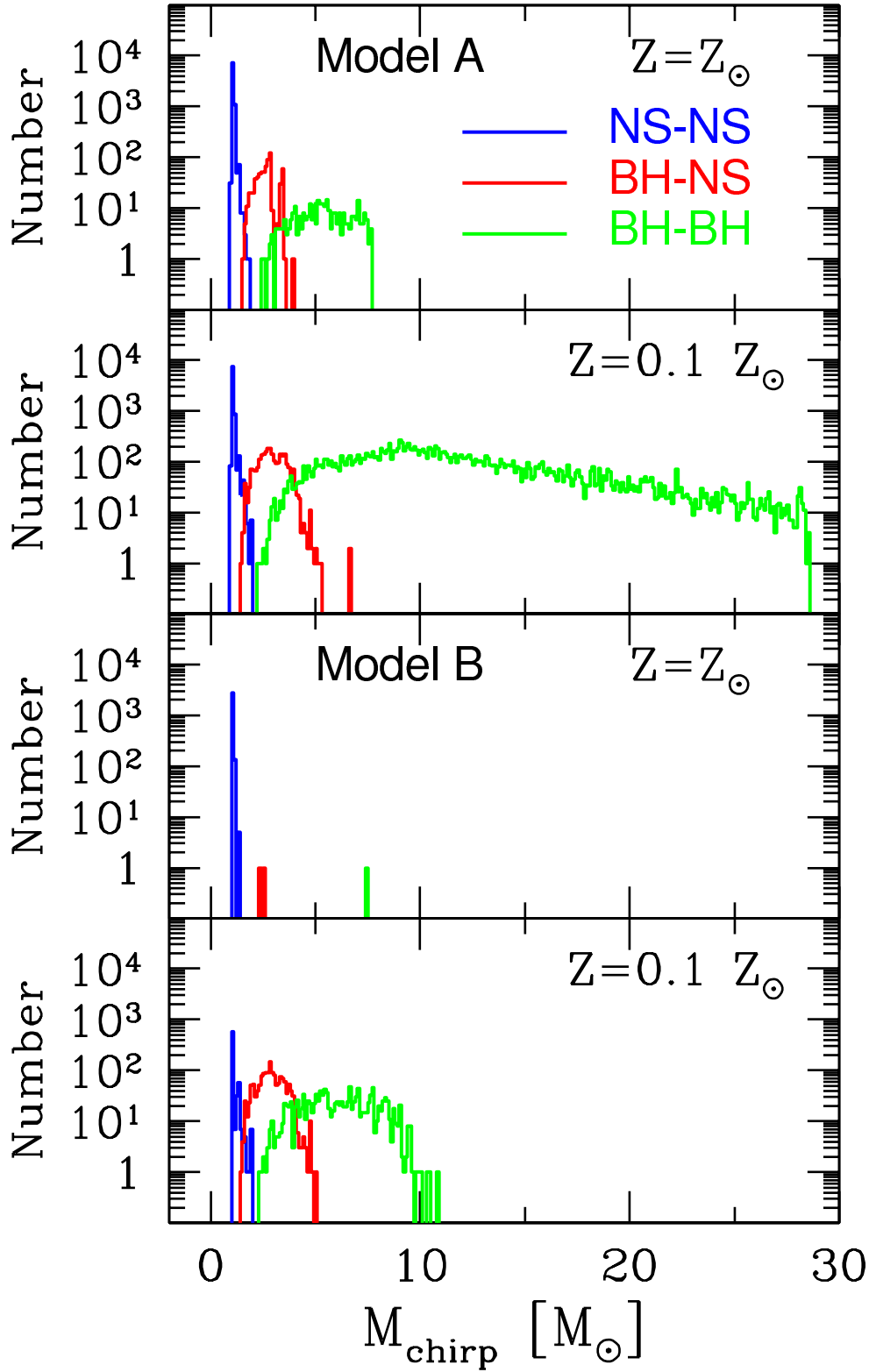}
\caption{
Chirp mass distribution for double compact objects. 
{\em Top two panels: Model A.} Note the strong effect of metallicity on chirp 
mass of binaries with black holes. Low metallicity (2nd panel down) reduces the wind mass
loss from the BH progenitors, allowing more massive BHs to form. 
The maximum chirp mass is $\sim 8 \msun$ for solar 
composition, while it can reach $\sim 30 \msun$ for $10\%$ solar for BH-BH 
mergers.
{\em Bottom two panels: Model B.} Note that BH binaries appear (in significant
numbers) only in the low metallicity case (bottom panel). The typical chirp mass
in model B is significantly lower than in model A. This is the result of progenitor 
elimination through 
common envelope mergers in model B. In particular, high mass stars (that can 
give birth to the highest mass BHs) reach large radii and are prone to enter 
a common envelope phase while crossing the Hertzsprung gap, thereby aborting further 
evolution even at low metallicity. 
}
\label{mc}
\end{figure}
\clearpage

\begin{figure}
\includegraphics[width=0.7\columnwidth]{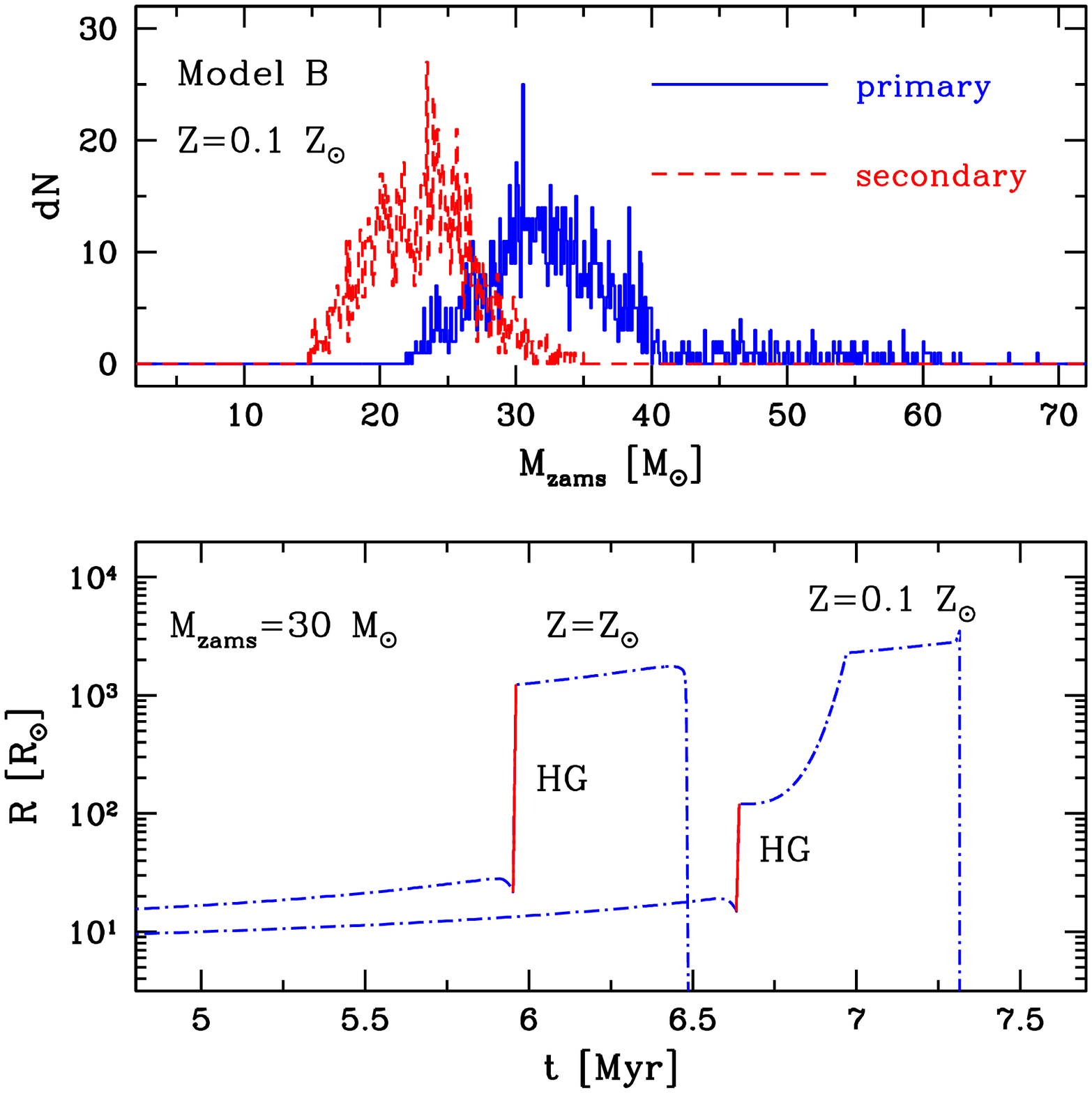}
\caption{
Explanation of the rate increase in Model B with decreasing metallicity.  
{\em Top panel.}
Initial masses (Zero Age Main Sequence) of stars that form black holes in
BH-BH binaries. Primaries (initially more massive components) transfer about 
$\sim 10 \msun$ in non-conservative Roche lobe overflow and rejuvenate the 
secondaries from their initial masses ($\sim 20\,\msun$) to $\sim 30\,\msun$. 
After the primary forms the first BH, the now more massive secondary evolves 
beyond Main Sequence and begins to rapidly increase its radius. 
{\em Bottom panel.}
The evolution of the radius of the $30 \msun$ star (rejuvenated secondary). 
In the Hertzsprung gap the radius increases to over $1000 \rsun$ for solar
metallicity, while for $10\%$ solar metallicity the radius reaches only $100 \rsun$. 
The rapid increase of the radius leads to a second Roche lobe overflow in the
progenitor. A common envelope ensues due to the high mass ratio of the
secondary to the primary BH. For solar metallicity 
the interaction is encountered while the secondary is in the Hertzsprung gap, and 
thus aborts formation of a BH-BH binary. For low metallicity the maximum 
radius during the Hertzsprung gap is relatively small, and some secondaries 
initiate the common envelope phase after having already become core-Helium burning 
stars. In such a case the system can survive the common envelope phase, and a close 
BH-BH binary is likely to form.
}
\label{hg1}
\end{figure}
\clearpage


\begin{references}

\reference{} Abadie, J., et al.\ 2010, CQG, submitted (arXiv:1003.2480) 
\reference{} Abbott, B., et al.\ 2008, ApJ, 683, L45
\reference{} Abbott, B., et al.\ 2009a, Nature, 460, 990
\reference{} Abbott, B., et al.\ 2009b, Phys. Rev. D., 80, 7101
\reference{} Belczynski, K., Kalogera, V., \& Bulik, T.\ 2002, ApJ, 572, 407
\reference{} Belczynski, K., et al.\ 2007, ApJ, 662, 504
\reference{} Belczynski, K., et al.\ 2008, ApJ Sup., 174, 223
\reference{} Belczynski K., et al.\ 2010a, ApJ, 714, 1217
\reference{} Belczynski, K., et al.\ 2010b, ApJ, 708, 117
\reference{} Belczynski, K., et al.\ 2010c, MNRAS, accepted (arXiv:0907.3486)
\reference{} Bulik, T., Belczynski, K., \& Prestwich, A.\ 2008, ApJ, 
             submitted (arXiv:0803.3516)
\reference{} Casares, J.\ 2007, IAU Symposium 238, 3
\reference{} Crowther, P., et al.\ 2010, MNRAS, 403, L41
\reference{} Dhawan, V., et al.\ 2007, ApJ, 668, 430
\reference{} Downing, J., et al.\ 2010, MNRAS, submitted (arXiv:0910.0546)
\reference{} Einstein, A.\ 1918, Preuss. Akad. Wiss. Berlin,
             Sitzungsberichte der physikalisch-mathematischen Klasse, 1, 154
\reference{} Flanagan, \'E.\'E. \& Hughes, S.A.\ 1998, PRD, 57, 4535
\reference{} Fryer, C., \& Kalogera, V.\ 2001, ApJ, 554, 548
\reference{} Hobbs, G., et al.\ 2005, MNRAS, 360, 974
\reference{} Hulse, R., \& Taylor, J.\ 1974, ApJ, 191, L59 
\reference{} Hurley, J., Pols, O., \& Tout, C.\ 2000, MNRAS, 315, 543
\reference{} Hut, P.\ 1981, A\&A, 99, 126
\reference{} Ivanova, N., \& Taam, R. E.\ 2004, ApJ, 601, 1058
\reference{} Martin, R., Tout, C., \& Pringle, J.\ 2010, MNRAS, 410, 1514  
\reference{} de Mink, S., et al.\ 2009, A\&A, 497, 243
\reference{} Mirabel, F., \& Rodrigues, I.\ 2003, Science, 300, 1119
\reference{} O'Shaughnessy, R., et al.\ 2008, ApJ, 672, 479
\reference{} Panter B., et al.\ 2008, MNRAS, 391, 1117
\reference{} Prestwich, A., et al.\ 2007, ApJ, 669, L21
\reference{} Sadowski, A., et al.\ 2008, ApJ, 676, 1162
\reference{} Taam, R. E., \& Sandquist, E. L. 2000, ARA\&A, 38, 113
\reference{} Timmes, F., Woosley, S., \& Weaver, T.\ 1996, ApJ, 457, 834
\reference{} Vazquez, G., et al.\ 2007, ApJ, 663, 995  
\reference{} Vink, J.S.\ 2008, New Astronomy Review, 52, 419
\reference{} Webbink, R.\ 1984, ApJ, 277, 355

\end{references}
\end{document}